# Tailoring the optical and physical properties of La doped ZnO nanostructured thin films

Mai M. A. Ahmed[1], Wael Z. Tawfik[1,2,*], M. A. K. Elfayoumi[1],

M. Abdel-Hafiez[3,4,5] and S. I. El-Dek[6,*]

[1]Department of Physics, Faculty of Science, Beni-Suef University, Beni-Suef, 62511, Egypt

[2]Department of Materials Science and Engineering, Chonnam National University, Gwangju 61186, Republic of Korea

[3]Lyman Laboratory of Physics, Harvard University, Cambridge, MA 02138, USA

[4]Physics Dept., Faculty of Science, Fayoum University, Fayoum, Egypt

[5]National University of Science and Technology "MISiS", Moscow 119049, Russia

[6]Materials Science and Nanotechnology Department, Faculty of Postgraduate Studies for Advanced Sciences (PSAS), Beni-Suef University, Beni-Suef, 62511, Egypt

**Abstract**

The modification and tailoring the characteristics of nanostructured materials are of great interest due to controllable and unusual inherent properties in such materials. A simple spray pyrolysis technique is used to prepare pure and La-doped ZnO films. The influence of La concentration (0, 0.33, 0.45, 0.66, 0.92 and 1.04 at. %) on the structural, optical, and magnetic properties of ZnO was investigated. The exact nominal compositions of the prepared films were determined from the field emission scanning electron microscope occupied with EDX. X-ray diffraction confirmed that the samples possessed single-phase hexagonal wurtzite structure. The main crystal size was decreased from 315.50 Å to 229.04 Å depending on La dopant concentration. This decrease is due to the small ionic radius of Zn ions in compared to La ions. The band gap values were found to be depend strongly on $La^{3+}$ ion content. Introducing La into ZnO induces a clear magnetic moment without any distortion in the geometrical symmetry, it also reveals the ferromagnetic coupling. The saturation magnetic moment of 1.04 at% La-doped ZnO shows the highest value of 0.014 emu, which is ~23 times higher than pure ZnO sample. The obtained results were discussed and compared with other literature data and showed an acceptable agreement.

[*] Corresponding authors

E-mail addresses: wael.farag@science.bsu.edu.eg (**Wael Z. Tawfik**), samaa@psas.bsu.edu.eg (**S. I. El-Dek**).

**1. Introduction**

The relation between semiconductors and magnetism has led to the next generation of magnetic semiconductors, where it is not the electron charge but the electron spin that carries information [1]. These diluted magnetic semiconductors (DMSs) are formed by the partial replacement of cations in a non-magnetic semiconductor by magnetic transition metal ions [2-5]. They are of keen potentials for various applications such as spintronics, spin-valve transistors, spin light-emitting diodes, and logic devices [6-7]. DMSs obtain their magnetic properties as a result of intrinsic defects or adding external impurities within semiconductors by doping process [6-8]. One of the most challenges facing DMS materials to be viable for commercial application is to get Curie temperatures ($T_c$) above room temperature or what so called the room temperature ferromagnetism (RTFM). Thus, recent studies are interested in increasing $T_c$ experimentally by controlling the different preparation conditions and through different doping mechanisms [8,9].

The state-of-the-art semiconductor materials those recorded as DMS materials are II-VI compounds including CdTe, HgTe, CdS, CdO, ZnS, and ZnO [10-14]. This is due to their ability to achieve many desirable properties such as morphological, optical and electronic. Earth-abundant metal oxide semiconductors with a large energy band gap have shown more efficient for DMS applications as it is easy in fabrication, availability and diversity in optical, chemical, electrical, magnetic and other characteristics. Furthermore, capability to adjust their properties by appropriate doping has made them very promising materials in a wide range of applications. Among all those earth-abundant metal oxide semiconductors, zinc oxide (ZnO) proved itself as a remarkable DMS material. ZnO presents strong solubility towards transition metals ions (3d) and rare earth ions (4f) which provides the needed **to** magnetic properties. DMS based on transition metals and rare earth doped ZnO are being studied as these impurity ions introduce ferromagnetic properties (intrinsic spin) of uncoupled electrons coming from unfilled 3d and 4f orbitals to be combined with free charge carriers in the same material [15,16]. Moreover, a non-toxic and inexpensive ZnO with a stable wurtzite structure doesn't only has a high transparency in the visible light region but it has also a direct wide band gap of 3.37 eV at room temperature which makes it one of the most encouraging materials for potential applications in various fields such as optoelectronic devices, transparent electrodes of solar cells, liquid crystal displays, memristors and gas sensing [15-19].

ZnO has been fabricated in different shapes and by different techniques. It was succeeded to be fabricated as platelets, clusters consisting of nano-crystals, nano wires and thin films as well [20-25]. Thin films are more practical and economical for different applications. ZnO thin film gives good adhesion towards different types of substrates. Previous studies have recorded variety of chemical and physical techniques for ZnO thin film deposition including chemical bath deposition, RF-sputtering, metal-organic chemical vapor deposition, molecular beam epitaxy, spray pyrolysis technique, filtered vacuum arc, and sol–gel [26-33]. Compared to other deposition techniques, spray pyrolysis is considered as a one of the most remarkable and promising thin film deposition techniques as it is distinguished in its ability to coat large-scale areas, simple, non-vacuumed technique and inexpensive [34,35].

Rare earth elements are also under focus as dopants in ZnO-based DMS materials because of their optical, electrical and magnetic properties [36,37]. Lanthanum (La)-doped ZnO nanoparticles prepared by sol-gel have given a room temperature ferromagnetism [38]. Recently, many researches are still working on showing the reasons of such behaviour. Some of those researches have interpreted such behaviour to oxygen vacancies in La-doped ZnO lattice [38]. Whereas other groups have backed this behaviour to Zn vacancies [15]. In addition, few theoretical studies showed that La-doped ZnO is diamagnetic material and impossible to be ferromagnetic [39]. That conflict is common to be noticed between the theoretical and experimental studies.

Thus, in this work a detailed study on the preparation of pure ZnO and La-doped ZnO films with different contents by simple spray pyrolysis technique is reported. It was found that the magnetic properties of ZnO film strongly depend on La content in ZnO matrix. By controlling the La content, different and controversial magnetic properties were reported along with structural, morphological and optical properties.

## 2. Materials and Methods

Chemicals used for growth pure ZnO and La-doped ZnO films were of analytical grade and used without any further purification. Zinc acetate dihydrate [Zn(CH$_3$COO)$_2$·2H$_2$O, 99%], deionized water (DIW), methanol, dihydrate lanthanum acetate [La(CH$_3$COO)$_3$·2H$_2$O, 99%] and glass were used as the precursor, solution, dopant and substrate respectively. The former aqueous solution for ZnO was dissolving 0.5 M dihydrated zinc acetate in 50 ml of mixed solvent DIW to absolute methanol of ratio 1:1. The pure ZnO sample was labeled as ZO-0. To obtain La-doped ZnO films, dihydrated lanthanum acetate is added to the precursor solution at different concentrations of 0.33, 0.45, 0.66, 0.92, and 1.04 at.% for samples LZO-1, LZO-2, LZO-3, LZO-4, and LZO-5, respectively. All aqueous solutions were stirred for 1 hr at room temperature for homogeneity. Prior to the deposition process, the glass substrates were firstly cleaned through immersing in diluted HCl solution with DIW followed by washing in 1:1 DIW to ethanol in order to remove any surface oxides. After that, the glass substrates were carefully cleaned in ultrasonic cleaner then dried properly using compressed air and transferred to the substrate holder (or heater). Throughout deposition of all the films, process parameters such as substrate temperature (450º ± 5 °C), air pressure (0.5 bar), solution flow rate (5 ml/min), distance between spray nozzle tip to the glass substrate (30 cm) and final solution concentration (0.5 M) were kept constant. In the interim, the spraying of the fine droplets of the prepared solution starts up to the heated substrate using compressed air as a carrying gas. The chemicals in the solution are vaporized react on the substrate surface after reaching it leading to the formation of La-doped ZnO (LZO) films on the glass substrates. The expected chemical reaction that could occur on the surface of the hot substrate is given by:

$$Zn(CH_3COO)_2 \cdot 2H_2O \xrightarrow{450°C} ZnO_{Film} + CO_2\uparrow + CH_3COCH_3\uparrow + H_2O\uparrow$$

X-ray diffraction (XRD) pattern were recorded using x-ray diffractometer (PANalytical

Empyr-ean) equipped with graphite monochromatized CuKα radiation (λ=1.54056 A˚, 40 kV, 35 mA) in the 2θ range = 20–70° to investigate the phase purity and structural parameters of the as-prepared films. JEOL-5410 field emission scanning electron microscopy (FE-SEM) attached with energy-dispersive X-ray (EDX) unit was employed to study the surface morphology and the elemental compositions of the as-prepared films. After photographing the films using FESEM, the micrographs were processed by Gwyddion 2.45 software without further calibrations [13]. After that, a 3D graph was initiated for each sample. The roughness parameters were calculated using the same software in nm. Optical transmittance spectra of all as-prepared films were measured in the wavelength range of 200–900 nm with double beam Perkin-Elmer UV-vis spectrophotometer at room temperature and the optical bandgaps were estimated from linear extrapolation of Tauc plots, assuming direct allowed transition (n = 2). The thicknesses of all films were determined from Parava software by counting the interference fringes formed on the samples. The magnetic moment measurements were performed using a superconducting quantum interference device magnetometer (MPMS-XL5) from Quantum Design.

## 3. Results and Discussion

Figure 1 represents structure examination using XRD for pure and La-doped ZnO films as collected at room temperature. It is clear that all films are formed in a polycrystalline structure coincided with hexagonal wurtzite phase. All patterns are matched well with ICDD card No. 01-075-1526 of space group P63mc no.186. Furthermore, there is no secondary phase referring to the complete solubility and homogeneity of $La^{+3}$ ion into ZnO lattice up to till ratio of 1.04 at.%. In addition, the main appeared peaks correspond to the planes (002), (101) and (103). It is obvious that peak of (002) has the largest intensity with respect to the others. This indicates that all films are preferred oriented along c-axis, which is perpendicular to substrate surface, and this behavior could be explained via Van der drift model [40]. It is also attributed to the nature of $sp^3$ hybrid orbitals of ZnO, which results in $Zn^{+2}$ and $O^{-2}$ ions tetrahedral coordination thereby lowering the inversing symmetry. It was found from previous studies that the lowest densities of the surface free energy were found to be 9.9, 12.3 and 20.9 $eV/nm^2$ that belongs to (002), (110) and (100), respectively [31]. Therefore, the film growth textured along (002) plane is mainly due to its lowest surface free energy [38]. These results agree well with that reported in references [27,31,33,38]. A significant plunged was observed in the main peak intensity (002) directly with rising $La^{+3}$ contents. This briefly is an indicator of lowering crystallinity as well as the ionic ordering in such orientation with increasing $La^{+3}$ on the expense of $Zn^{+2}$ cations in equivalent lattice sites. Such behaviour may be a reflection of micro-strain arises due to the significant difference in ionic radii and in charge imbalance between $Zn^{+2}$ and $La^{+3}$ [27,33]. This micro-strain is more pronounced with further replacing $Zn^{+2}$ by $La^{+3}$ and in our case without altering the solubility of the latter into the hexagonal matrix of ZnO. It is noticed that the values of micro-strain, as calculated from (002) plane, achieved around 1.8 times of its value at La content of 1.04 at.% when compared to the un-doped parent ZnO. On the other hand, Y. Babacan *et al.* and Y. Bouznit *et al.* [19,31] recorded an increase of (002) peak intensity regarding to La content. Broadening and shift in the main peak intensity demonstrated successful incorporation of $La^{+3}$ ions into ZnO crystal lattice. When $La^{+3}$ ion occupy $Zn^{+2}$ interstitials sites, which this causes lattice distortion and expansion because of its larger ionic radii of $La^{+3}$ (1.032 Å) as compared to that of $Zn^{+2}$ (0.734 Å). Crystallite size was calculated using Debye-Scherrer equation [27]:

$$D = \frac{K\lambda}{\beta \cos\theta} \qquad (1)$$

Where k= 0.9 represent shape factor, λ= 1.54056 Å the wavelength of CuK$_\alpha$ incident beam, β= corrected line broadening at half maximum in radians and θ is Bragg angle. According to the calculated values reported in table (1), there is a reduction in the crystallite size against La$^{+3}$. Depending on similar results reported by S. Anandan *et al.* [41] using co-precipitation method, I. Stambolova *et al.* [42] using spray pyrolysis technique and W. Lan *et al.* [27] using R.F magnetron sputtering encountered the similar behavior as well. This behavior was mainly attributed to La$^{+3}$ incorporation within ZnO lattice sites may obstruct the grain growth and limit the grain boundary movement. Moreover, it may due to formation of Zn-La-O on doped thin film surface. This will lurk the grain growth and reduces the crystallite size. Another interpretation introduced by *Eyup et al.* [33] using sol-gel technique is that when large amount of La$^{+3}$ content of ionic radius 1.032 Å replaces Zn$^{+2}$ of ionic radius 0.734 Å inside ZnO lattice; a lattice distortion could be induced where crystal defects occurred. As a result, an impediment in the film growth could be formed, which could diminish the lattice size. This reduction in the grain size is a common behaviour in rare earth doped ZnO [36,37]. This could be attributed to the similarity in their ionic radius.

The unit cell dimensions a and c along c-axis were estimated according to hexagonal wurtzite structure and listed in table (1) by using the equation [38]:

$$\frac{1}{d_{hkl}^2} = \frac{4}{3}\left(\frac{h^2 + hk + k^2}{a^2}\right) + \left(\frac{l^2}{c^2}\right) \qquad (2)$$

where d is the inter planar distance (hkl) are the Miller indices for the main peaks. Zn–O bond length of was also calculated using the following relation [43]:

$$L = \sqrt{\frac{a^2}{3} + (\frac{1}{2} - u)^2 c^2} \qquad (3)$$

where parameter u is represented by:

$$u = \frac{1}{2} + \frac{a^2}{3c^2}$$

where a and c are the lattice parameter calculated above. The volume of the unit cell was estimated by using equation [27]:

$$V = \frac{\sqrt{3}}{2} a^2 c \qquad (4)$$

As listed in table (1) there is a slight increase in lattice constant, bond length and lattice volume reflecting size difference between cations, this may agree with the fact that, there is significant difference between $La^{+3}$ to $Zn^{+2}$ radii. As $La^{+3}$ substitutions into lattice occur, expansion was expected. Therefore, the Zn–O bond length is increased when some $La^{+3}$ ions reside in the unit cell at $Zn^{+2}$ ion positions. The theoretical density was computed using [12]:

$$D_x = \frac{ZM}{NV} \qquad (5)$$

where Z is the number of molecular per unit cell (Z=2), as depicted from the ICDD card, M is the molecular weight of the investigated samples, N is the known Avogadro's number and V is the unit cell volume. This likely seen that the density depends on both molecular weight and volume of unit cell. As the difference between atomic weights of $La^{+3}$ (138.91 amu) and $Zn^{+2}$ (65.37 amu) is almost the double, then it will compensate the slight increase in the unit cell volume. This in turns explained the observed enlargement of $D_x$ values from 5.23 g/cm$^3$ for pure ZnO to 5.54 g/cm$^3$ for the La rich sample (1.04 at.%).

Figure 2 (a:f) displays FE-SEM micrographs of the as-prepared thin films of La:ZnO at different contents of La. It is obvious that the pure sample was formed as a cracked surface. However, with rising La ions into the films, grains tend to be more spherical. In addition, surface is distinguished with inter-granular porosity. The ratio of this porosity increases with La content growing. In addition, in the high contents' samples of La doping, the grains tend to be denser and compact, the micro cracks disappear and the grains look to be more observed and defects are more pronounced. These defects are generally due to the difference between the thermal expansion of both ZnO thin film ($\alpha = 7 \times 10^{-6}/°c$) and soda lime glass ($\alpha = 9 \times 10^{-6}/°c$), besides the lattice mismatch between glass substrate and perfectly order hexagonal ZnO [31].

Roughness investigation of thin films is illustrated in Fig. 3. The scan area was taken to be 91×59 μm$^2$, 3-dimensional image. The obtained values of root mean square (RMS) roughness are listed in table (1). It is observed that, RMS roughness tends to rise with La content increasing which may refer to the great influence of La ions into ZnO grains. It started with 27 nm for pure ZnO sample, then it grew steadily to achieve around 59.5 nm, and then it deteriorated suddenly to be 47.3 nm at the highest content of La. This trend of RMS could be explained by supposing that La ions have been trapped successfully into ZnO lattice, therefore the former work as a source of lattice defects [13]. Consequently, rising of La content leads to growing of surface roughness. This rough of surface may induce interactions with host materials which may promote this type of thin film for gas sensing applications [18]. The energy-dispersive X-ray (EDS) analysis spectrum of pure and 1.04 at.% La-doped ZnO are shown in Fig. 4 a and b. EDS analysis unambiguously confirms the presence of La in the ZnO deposited films. The average atomic percent of La, Zn and O elements are reported and listed in table (1). Figure 4c shows the EDS elemental mapping of the 1.04 at.% La-doped ZnO sample. This analysis shows the distribution of La, Zn and O ions into the lattice surface.

Optical transmittance spectra of pure ZnO and La-doped ZnO films with different La contents are displayed in Fig. 5a. It is observed that all films exhibited high transparency in the visible region of 400–800 nm. Furthermore, the appearance of distinct Fabry–Perot fringes in the transmittance spectra of all deposited films are solid evidence of the high quality and homogeneity of the films surface which in turn manifest that the films are uniform and smooth. The thicknesses of all sprayed films were calculated using the interference method through PARAV software and results are listed in table (1). The optical band gaps (Eg) of all films were estimated from the Tauc's relationship between the absorption coefficient $(\alpha)^2$ and the photon energy (hν) using the given equation [44, 45]:

$$\alpha h\nu = A(h\nu - E_g)^n \quad (6)$$

where A is the slope of Tauc line which is knowing as a band tailing parameter and n values depending on the type of optical transition, for direct allowed optical band gap n = 2. Using the well-known relation between $(\alpha h\nu)^2$ *vs.* hν, the Eg values for all films were determined by the extrapolation method as illustrated in Fig. 5b. The obtained values of Eg are listed in table (1). As expected from the absorption edges, the Eg showed two different behaviors against La contents. The Eg values were increased from 3.214 eV to 3.245 eV as La content was increased from 0 at.% to 0.45 at%. Then it starts to descend upto 3.194 eV with a further increase in La content upto 1.04 at.%, respectively.

Band-gap widening with increasing La content in ZnO matrix is believed to be due to the Burstein-Moss (B-M) shift [46] by the higher electron density in the films through replacing $Zn^{2+}$ ions by $La^{3+}$ ions. Pure ZnO sample has no free carrier charges due to the ionic bond between $Zn^{2+}$ and $O^{2-}$. With replacing $Zn^{2+}$ ions by $La^{3+}$ ions in ZnO matrix there will be extra free electrons in the valence band. Since the lowest states in the conduction band is fulfilled by electrons so that an additional energy is required for electrons in valence band to jump into the empty states in the conduction band. Based on the above explanation, one can expected that the values of $E_g$ will keep increasing with increasing $La^{3+}$ content due to the higher electron density in the films. Nevertheless, it was observed that the $E_g$ values were decreased at high La content (La $\geq$ 0.45 at.%) which is in slightly conflict with B-M effect. This could be attributed to the formation of crystal defects in ZnO matrix with increasing La content due to the large difference in the ionic radii between $Zn^{2+}$ and $La^{3+}$ ions. At high La content, the interstitials occupation of $Zn^{2+}$ ions by $La^{3+}$ ions lead to a lattice distortion and expansion which in turn gives a new level of localized states around the conduction band causing additional transitions of free electrons in valence band to the defects band rather than the conduction band itself. The results are consistent with the XRD data.

Figure 6a presents the temperature dependence of the magnetic moment measured up to 300 K with the magnetic field of 1T. It is clear that for pure ZnO sample, the paramagnetic behavior predominates which is well known for the pure ZnO as shown in Fig. 6b. La-doped ZnO exhibits a clearly magnetic hysteresis loop. It can be seen that there is an enhancement of the magnetic moment by increasing La doping in contrast to pure ZnO sample as revealed through the moment values. Theoretically Chu *et al.* proof that ferromagnetism could be induced by the exchange interaction between La ions and Zn ion spin moments [47]. One can see that there is an enhancement of the magnetic moment by increasing doping in contrast to pure ZnO sample as revealed through the magnetization. It is apparent that there is a proper long-range ferromagnetic order in the rich doped sample which is sharper than the lower doping. The reason can also be seen in the hysteresis curve shown in the Fig. 6b. The weak ferromagnetism in our investigated samples is clear from the lack of even at higher fields. La ions substituting into the ZnO induced weak ferromagnetism and illustrates a magnetic moment without any distortion in the geometrical symmetry. It is important to note that the reduced moment values might occur due to large disordered spins in the surface as well due to grain boundaries that constitute a considerable fraction of the nanosample. By increasing $La^{3+}$ content in the samples, we expected that some vacancies are generated due to the difference in the valence state between $Zn^{2+}$ and $La^{3+}$. These vacancies will be ordered in the unique axis of magnetization (c-axis) in our case as the films exhibited hexagonal symmetry. The magnetic moment of 1.04 at% ($14 \times 10^{-3}$ emu) film revealed 23 times higher than that of the pure ZnO one ($0.6 \times 10^{-3}$ emu). Finally, the diamagnetic dilution in the hexagonal diamagnetic lattice films of wurtzite structutre resulted in improved values of magnetic moment and ferromagnetic character which needed further investigations in our future work [48-52].

## 4. Conclusion

Pure and La doped ZnO thin films were successfully synthesized in single phase hexagonal structure belonging to the space group P63mc. The unit cell volume decreased with poor values and then increased again. La doping enhanced the densification but impeded the grain growth. The microstrain values are more pronounced with La doping content. The band gap increased with La doping while for high La contents and then it decreased again. The pure ZnO revealed paramagnetic trend, whilst the La doping induced ferromagnetism at different levels. From our point of view, one could recommend these films for magnetic gas sensors and in spintronic applications.

**Acknowledgment:**

This research was supported by the Academy of Scientific Research and Technology through the Scientists for Next Generation (SNG-2015) Program (Grant No. FRM-SGO-22). MA thanks the Ministry of Education and Science of Russia through NUST (MISiS) Grant No. K2-2017-084 and the Act 211 of the Government of Russia, contract 02.A03.21.0004 and 02.A03.21.0011.

[23] S. A. Ansari, M. M. Khan, M. O. Ansari, J. Lee, M. H. Cho, Biogenic Synthesis, Photocatalytic, and Photoelectrochemical Performance of Ag−ZnO Nanocomposite, J. Phys. Chem. C 117 (2013) 27023−27030.

[24] H.-Y. He, J.-F. Huang, J. Fei, J. Lu, La-doping content effect on the optical and electrical properties of La-doped ZnO thin film, J. Mater. Sci. Mater. Electron. 26 (2014)1205-1211.

[25] M. M. Khana, N. H. Saadah, M. E. Khan, M. H. Harunsani, A. L. Tan, M. H. Cho, Potentials of Costus woodsonii leaf extract in producing narrow band gap ZnO nanoparticles, Mater. Sci. Semicond. Process. 91 (2019) 194–200.

[26] S. K.Shaikh, S. I. Inamdar, V. V. Ganbavle, K. Y. Rajpure, Chemical bath deposited ZnO thin film based UV photoconductive detector, J. Alloys Compd. 664 (2016) 242-249.

[27] W. Lan, Y. Liu, M. Zhang, B. Wang, H. Yan, Y. Wang, Structural and optical properties of La-doped ZnO films prepared by magnetron sputtering, Mater. Lett. 61 (2007) 2262–2265.

[28] M. A. Hassan, M. A. Johar, S. Y. Yu, S.-W. Ryu, Facile Synthesis of Well-Aligned ZnO Nanowires on Various Substrates by MOCVD for Enhanced Photoelectrochemical Water-Splitting Performance, ACS Sustainable Chem. Eng. 6 (2018) 16047−16054.

[29] Y. W. Heo, K. Ip, S. J. Pearton, D. P. Norton, J. D. Budai, Growth of ZnO thin films on c-plane $Al_2O_3$ by molecular beam epitaxy using ozone as an oxygen source, Appl. Surf. Sci. 252 (2006) 7442–7448.

[30] F. Gül, H. Efeoglu, Bipolar resistive switching and conduction mechanism of an Al/ZnO/Al-based memristor, Superlattices Microstruct. 101 (2017) 172-179.

[31] Y. Bouznit, Y. Beggah, F. Ynineb, Sprayed lanthanum doped zinc oxide thin films, Appl. Surf. Sci. 258 (2012) 2967-2971.
17

**Tables**

**Table 1.** Values of XRD, Roughness, EDX, optical and magnetic parameters for all as-prepared films.

| Sample | ZO-0 | LZO-1 | LZO-2 | LZO-3 | LZO-4 | LZO-5 |
|---|---|---|---|---|---|---|

| Lattice parameter, a (Å) | 2.9135 | 2.9133 | 2.9130 | 2.9123 | 2.9142 | 2.9159 |
|---|---|---|---|---|---|---|
| Lattice parameter, c (Å) | 4.3702 | 4.3699 | 4.3699 | 4.3685 | 4.3713 | 4.3739 |
| Unit cell volume, V (Å)$^3$ | 36.999 | 36.974 | 36.959 | 36.938 | 37.032 | 37.086 |
| Theoretical density, $D_x$ (g/cm$^3$) | 12.676 | 12.684 | 12.683 | 12.698 | 12.706 | 12.659 |
| Crystal size (Å) | 315.50 | 251.57 | 229.04 | 229.77 | 263.12 | 274.44 |
| Micro-strain (10$^{-3}$ %) | 5.5 | 6.45 | 6.28 | 6.77 | 6.65 | 9.8 |
| Roughness average, Ra (nm) | 20.9 | 31.5 | 40.8 | 44.5 | 46.6 | 37.2 |
| Root mean square roughness, Rq (nm) | 27.0 | 39.3 | 51.1 | 57.0 | 59.5 | 47.3 |
| Atomic % — Zn K | 54.15 | 51.97 | 50.72 | 51.83 | 54.18 | 56.85 |
| Atomic % — O K | 45.85 | 47.69 | 48.83 | 47.51 | 44.91 | 42.11 |
| Atomic % — La L | 0.00 | 0.33 | 0.45 | 0.66 | 0.92 | 1.04 |
| Film thickness (μm) | 1.284 | 0.788 | 1.212 | 0.684 | 0.511 | 0.566 |
| Energy Bandgap, $E_g$ (eV) | 3.214 | 3.222 | 3.245 | 3.236 | 3.205 | 3.194 |
| Saturation magnetic moment, $M_s$ (10$^{-3}$.emu) | 0.63 | -- | -- | 9.69 | 12.34 | 14.38 |

**Figure Captions**

Fig. 1. XRD pattern of La-doped ZnO at different atomic ratios ranging from 0 to 1.04%.

Fig. 2. FE-SEM images of pure ZnO and La-doped ZnO films.

Fig. 3. Surface roughness of La-doped ZnO films as a function of La content.

Fig. 4. EDS spectrum of (a) pure ZnO, (b) 1.04 at.% La-doped ZnO film and (c) EDS elemental mapping of 1.04 at.% La-doped ZnO sample (LZO-5).

Fig. 5. (a) Optical transmittance spectra against wavelength for pure and La-doped ZnO films and (b) the plots of $(\alpha h\nu)^2$ vs. $h\nu$ and optical band-gaps evaluated by extrapolation method.

Fig. 6. (a) Magnetic moment vs. temperature of the two investigated samples under a magnetic field of 1T and (b) illustrates the magnetic hysteresis (M-H) curves of three La doped samples measured at 2K.

Fig. 1

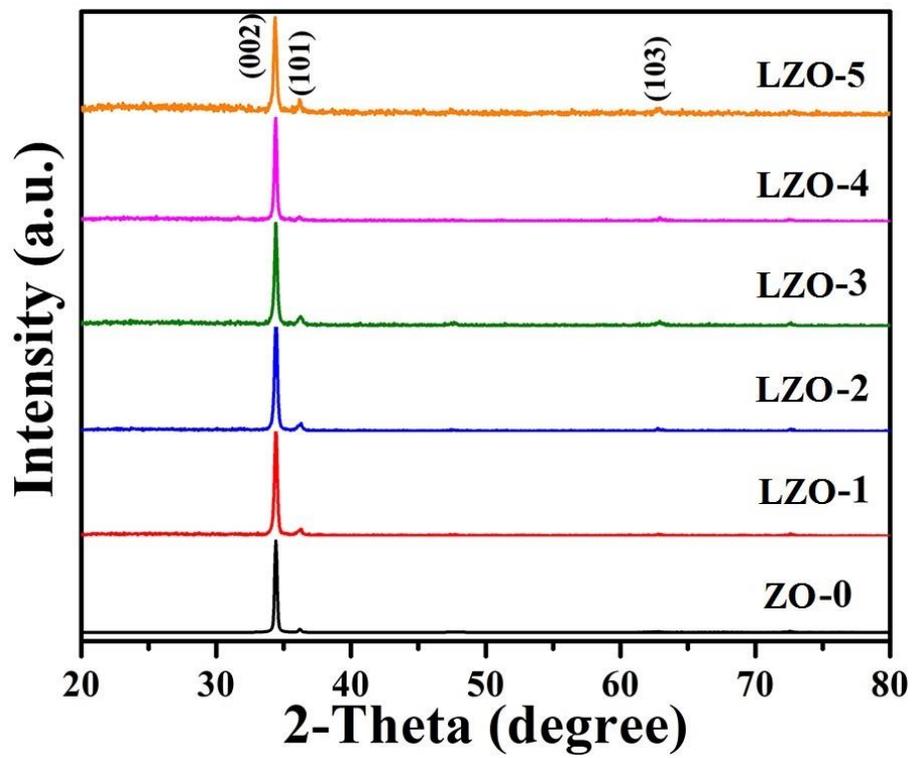

Fig. 2

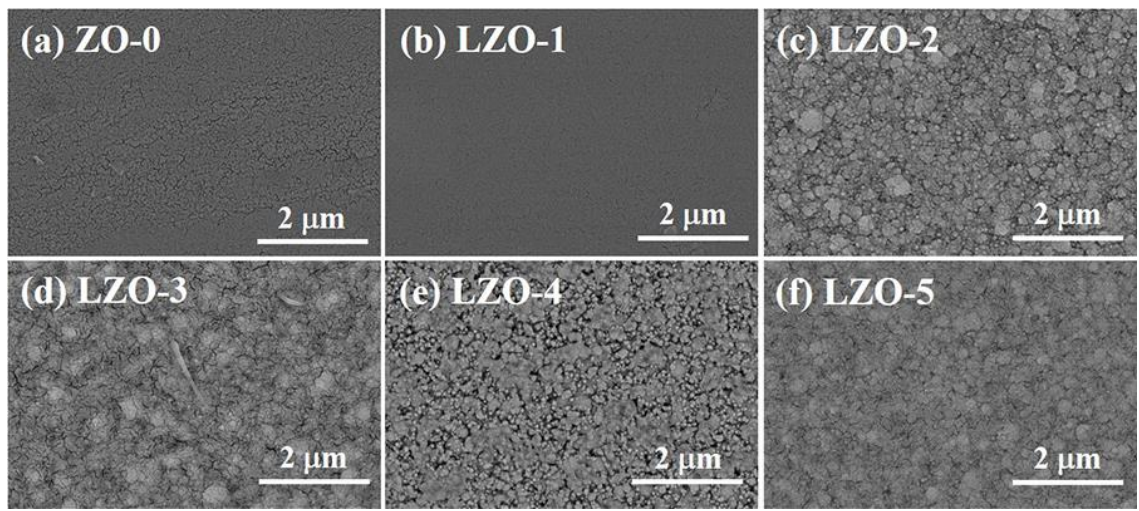

Fig. 3

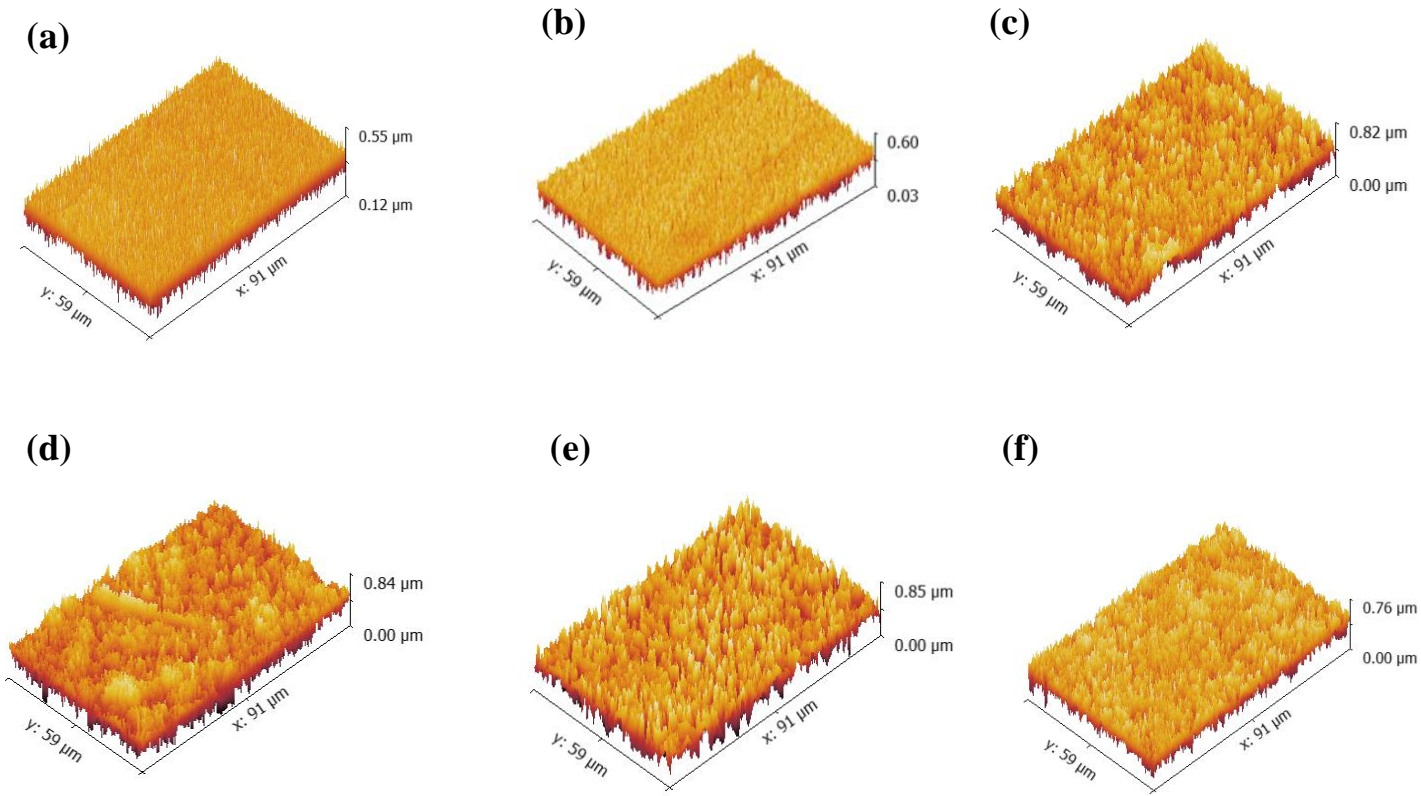

Fig. 4

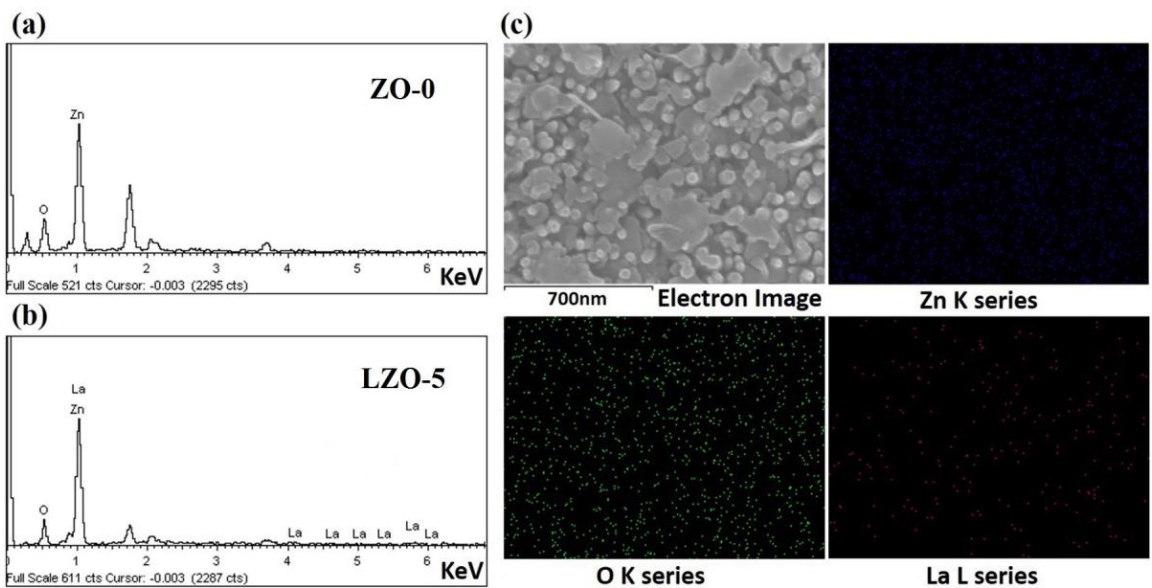

Fig. 5

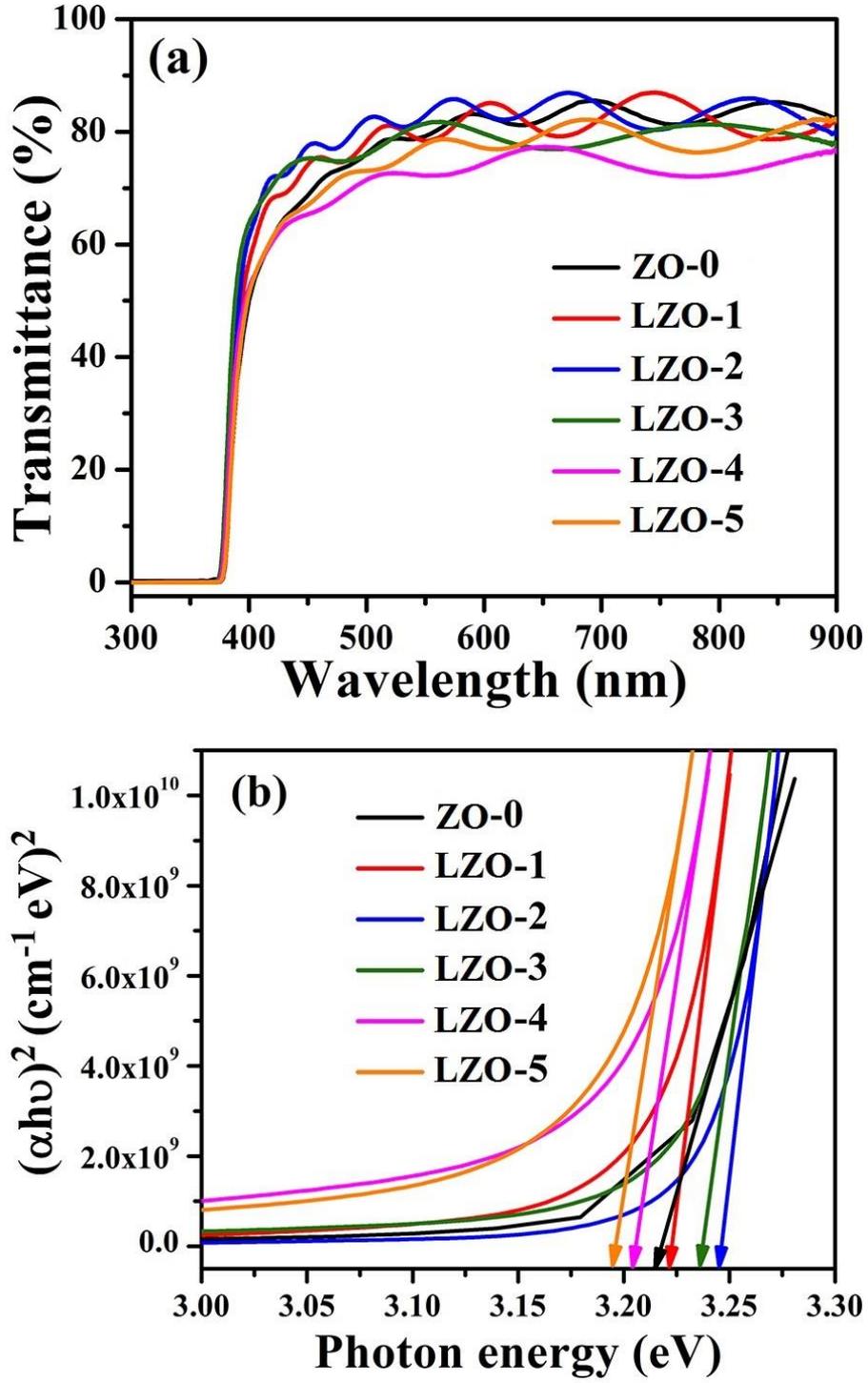

Fig. 6

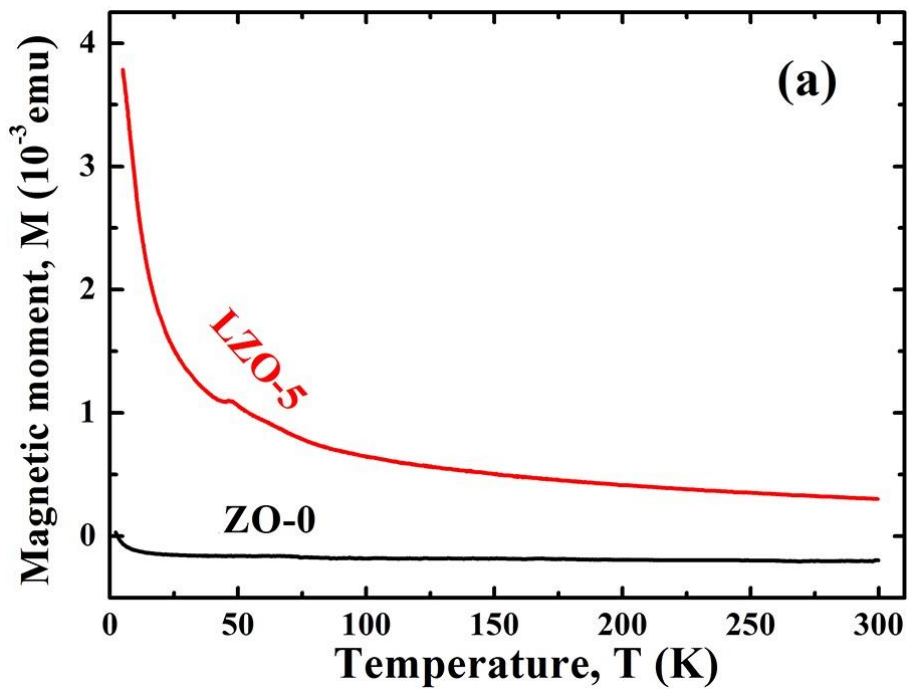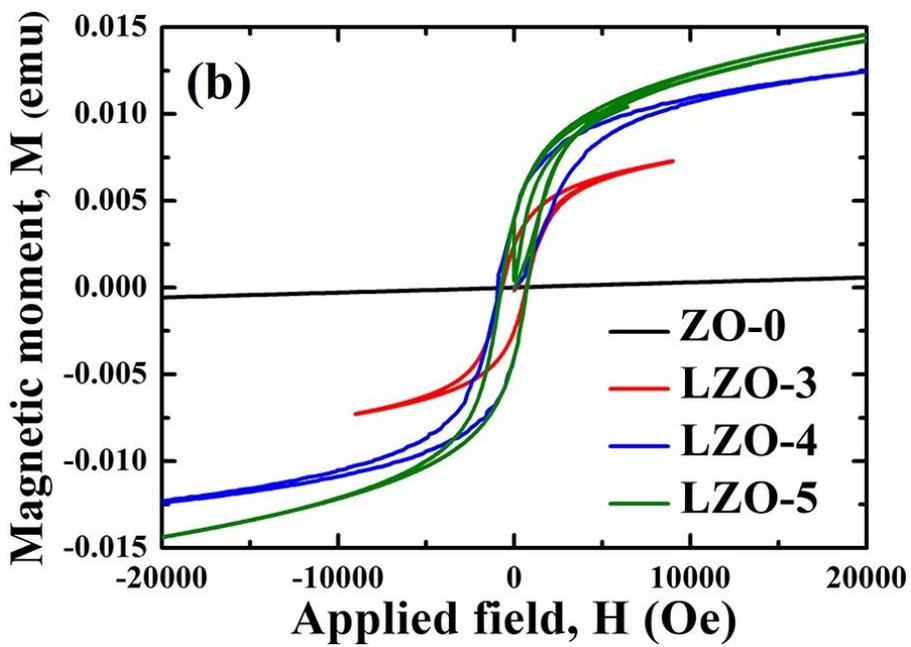